\documentclass[%
 aps,      
 pre,      
 amsmath,amssymb,
 reprint,%
 groupedaddress,
 showpacs,
]{revtex4-1}

\usepackage{graphicx}
\usepackage{dcolumn}
\usepackage{bm}
\usepackage{verbatim}
\usepackage{color}

\begin{document}

\title{One-Dimensional Model of Inertial Pumping}

\author{Pavel E. Kornilovitch}
 \email{pavel.kornilovich@hp.com}
 \affiliation{Hewlett-Packard Company, Printing and Personal Systems, Corvallis, Oregon 97330, USA} 

\author{Alexander N. Govyadinov}
 \affiliation{Hewlett-Packard Company, Printing and Personal Systems, Corvallis, Oregon 97330, USA} 
 
\author{David P. Markel}
 \affiliation{Hewlett-Packard Company, Printing and Personal Systems, Corvallis, Oregon 97330, USA}  

\author{Erik D. Torniainen}
 \affiliation{Hewlett-Packard Company, Printing and Personal Systems, Corvallis, Oregon 97330, USA} 

\date{\today}  

\begin{abstract}

A one-dimensional model of inertial pumping is introduced and solved. The pump is driven by a high-pressure vapor bubble generated by a microheater positioned asymmetrically in a microchannel. The bubble is approximated as a short-term impulse delivered to the two fluidic columns inside the channel. Fluid dynamics is described by a Newton-like equation with a variable mass, but without the mass derivative term. Because of smaller inertia, the short column refills the channel faster and accumulates a larger mechanical momentum. After bubble collapse the total fluid momentum is nonzero, resulting in a net flow. Two different versions of the model are analyzed in detail, analytically and numerically. In the symmetrical model, the pressure at the channel-reservoir connection plane is assumed constant, whereas in the asymmetrical model it is reduced by a Bernoulli term. For low and intermediate vapor bubble pressures, both models predict the existence of an optimal microheater location. The predicted net flow in the asymmetrical model is smaller by a factor of about 2. For unphysically large vapor pressures, the asymmetrical model predicts saturation of the effect, while in the symmetrical model net flow increases indefinitely. Pumping is reduced by nonzero viscosity, but to a different degree depending on the microheater location.        

\end{abstract}


\maketitle

\section{\label{oned:sec:one}
Introduction
}

In a recent paper~\cite{Torniainen2012} we reported numerical modeling and experimental demonstration of inertial pumping. In this effect~\cite{Yuan1999b,Yin2005a,Yin2005b}, fluid is confined within a thin channel connecting two large reservoirs and pumped by a high-pressure vapor bubble that repeatedly expands and collapses. The bubbles can be created by a number of actuation methods including thermal resistors~\cite{Yin2005b,Torniainen2012}, electrical currents passing through the fluid~\cite{Geng2001,Yin2005a}, localized laser pulses~\cite{Wang2004,Dijkink2008,Sun2009}, and acoustic actuation~\cite{Dijkink2006}. If the bubbles are generated away from the geometric center of the channel, the dynamics of the two fluidic columns are asymmetric, which results in a net flow from the short toward the long arm of the channel.    

The pumping action has been attributed to unequal inertia of the two fluidic columns inside the channel~\cite{Yuan1999b,Yin2005b,Dijkink2006,Torniainen2012}. While several numerical studies have been done in realistic three-dimensional geometries~\cite{Ory2000,Sun2009,Torniainen2012}, the main effect can be captured within a simple one-dimensional model derived from momentum balance~\cite{Yuan1999b,Yin2005a,Yin2005b,Sun2009,Torniainen2012}. The model has been used to prove nonzero net flow~\cite{Yin2005b} and illustrate the main qualitative features of the pump through numerical analysis~\cite{Torniainen2012}. 

The model is nonlinear and in general not solvable analytically. However, for realistic values of pressure (several atmospheres), channel width (tens of microns), and fluid properties (density, viscosity, and surface tension of water), the dynamics is dominated by the balance between inertia and pressure forces~\cite{Torniainen2012}. If only those two factors are left in the dynamic equation, the latter simplifies so much that an analytical solution becomes possible. The goal of this paper is to present this solution, which provides insight into the mechanism of inertial pumping and becomes a useful tool for studying this effect.

\section{\label{oned:sec:two}
The model
}

\begin{figure}[t]
\includegraphics[width=0.48\textwidth]{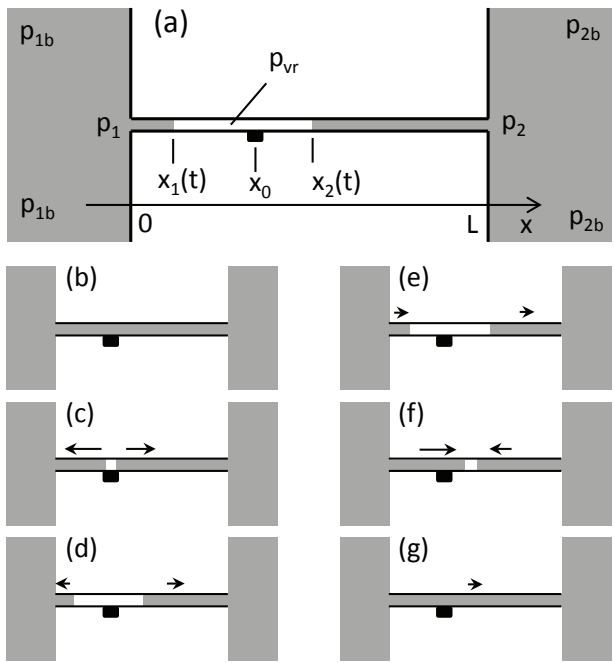}
\caption{(a) Schematic geometry of an inertial micropump. (b)--(g) Phases of the expansion-collapse cycle. (b) In the starting state, the fluid is at rest. (c) The microheater (black square) creates a high-pressure vapor bubble. A positive pressure difference pushes the fluid out of the channel. (d) The vapor bubble pressure quickly drops below atmospheric and the fluid decelerates under a negative pressure difference while moving by inertia. (e) The short arm turns around, while the long arm is still moving out of the channel. (f) The two fluidic columns collide at a point shifted from the starting point of the expansion: the primary pumping effect. (g) Since the shorter arm has a larger momentum at collision, the total postcollapse momentum is nonzero: the secondary pumping effect.}   
\label{oned:fig:one}
\end{figure}

Consider the geometry of Fig.~\ref{oned:fig:one}(a). A thin channel of cross-sectional area $A$ is connected to two much wider reservoirs. The reservoirs and the channel are filled with incompressible fluid of density $\rho$. The pressures in the reservoirs far from the channel are $p_{1b}$ and $p_{2b}$, respectively. In the most common case, both bulk pressures are equal to the atmospheric pressure $p_0$, but more general cases may be considered. Inside the channel is a resistive microheater that locally boils the fluid and creates high-pressure vapor bubbles. Each bubble expands and collapses, resulting in a nonzero net flow through the channel [see Figs.~\ref{oned:fig:one}(b)-\ref{oned:fig:one}(g)]. The microheater must be positioned asymmetrically relative to the channel's ends to achieve pumping. 

The simplest approach to flow dynamics is to completely neglect transverse motion of the fluid in the channel, three-dimensional features of the bubble shape, and curvature of the vapor-fluid interface. These assumptions are approximate; however, they do not change the qualitative picture of inertial pumping. Through full three-dimensional computational fluid dynamics (CFD) modeling we showed that pumping is most efficient when the bubble occupies the entire cross section of the channel pushing the fluid primarily along the channel's axis~\cite{Torniainen2012}. Under these assumptions, the channel is approximated by a one-dimensional line and a vapor-fluid interface by a single point on this line. Thus a single bubble is fully described by the time-dependent positions of two interfaces $x_1$ and $x_2$. The case of more than one bubble can be formulated in the same manner. The goal of the present paper is to analyze the single bubble case.

\subsection{\label{oned:sec:twoone}
Momentum balance and forces
}

Consider the left column of fluid (see Fig.~\ref{oned:fig:one}). Its length is $x_1(t)$, velocity $\dot{x}_1$, and total momentum $Q_1(t) = \rho A \, x_1 \, \dot{x}_1$. In a time interval $dt$ the momentum changes because of two factors: (i) the force $F_1(t)$ acting on the fluid and (ii) momentum lost to or supplied by the left reservoir. During the expansion phase $\dot{x}_1 < 0$ and the column loses a negative momentum $\rho A \dot{x}^2_1 \, dt$ to the reservoir. This element must be added to momentum balance as a positive increment:
\begin{equation}
Q_1(t + dt) = Q_1(t) + \rho A \dot{x}^2_1 \, dt + F_1(t) \, dt \: .
\label{onedf:eq:one}
\end{equation}
During the collapse phase $\dot{x}_1 > 0$, and the column acquires a positive momentum $\rho A \dot{x}^2_1 \, dt$ from the reservoir and the same equation applies. Substituting $Q_1$ one derives the dynamic equation~\cite{Yuan1999b,Yin2005a,Yin2005b,Torniainen2012}  
\begin{equation}
\rho A \, x_1 \, \ddot{x}_1 = F_1(t) \: .
\label{onedf:eq:two}
\end{equation}
This is a variable-mass Newton equation, but without the mass-derivative term: The latter has been dissipated by the reservoir. A similar momentum balance analysis yields, for the right fluidic column,
\begin{equation}
\rho A \, ( L - x_2 ) \, \ddot{x}_2 = F_2(t) \: ,
\label{onedf:eq:three}
\end{equation}
where $L$ is the total channel length. In this case, the right reservoir absorbs or supplies the extra mechanical momentum.

The major forces at play during the expansion-collapse cycle are (i) pressure-difference forces, (ii) viscous forces, and (iii) surface tension forces. The surface tension forces are about two orders of magnitude smaller than the other two for typical microfluidic conditions~\cite{Torniainen2012} and can be safely neglected. The viscous force is typically smaller than the pressure force, but is not always negligible. The simplest way to account for the viscous force is to assume that it is proportional to velocity and also to the length of the column because it acts along the fluid-surface boundary. The resulting one-dimensional model reads    
\begin{eqnarray}
A \rho \,       x_1   \, \ddot{x}_1 + \kappa \, x_1 \,         \dot{x}_1 & = & ( p_1 - p_{\rm v} ) A \: ,
\label{onedf:eq:four} \\
A \rho \, ( L - x_2 ) \, \ddot{x}_2 + \kappa \, ( L - x_2 ) \, \dot{x}_2 & = & ( p_{\rm v} - p_2 ) A \: .
\label{onedf:eq:five}
\end{eqnarray}
Here $\kappa$ characterizes the relative strength of the viscous force. Under further assumptions, $\kappa$ can be linked to other parameters of the system. For example, within the Poiseuille flow
\begin{equation}
\kappa = 8 \pi \eta \: ,
\label{onedf:eq:six}
\end{equation}
where $\eta$ is the bulk viscosity of the fluid. This relationship is derived in the Appendix. In this work $\kappa$ will be treated as an independent phenomenological parameter.

\subsection{\label{oned:sec:twotwo}
Impulse bubbles and boundary conditions
}

The next step is to specify the pressures in Eqs.~(\ref{onedf:eq:four}) and (\ref{onedf:eq:five}). There $p_{\rm v}$ is the vapor pressure inside the vapor bubble, while $p_{1,2}$ are the fluid pressures inside the reservoirs near the channel's ends [see Fig.~\ref{oned:fig:one}(a)]. All pressures are time dependent and require separate considerations. Let us begin with the vapor bubble. 

A typical pumping event starts with a microheater boiling the surrounding fluid within a fraction of a microsecond. This creates a thin layer of superheated vapor with pressure of about 8--10 atm \cite{Zhao2000}. This high-pressure phase lasts about 1 $\mu$s. Then the pressure quickly drops to subatmospheric levels because of bubble expansion and heat transfer losses. The residual bubble pressure (the saturation pressure of the vapor at the ambient temperature) is about $p_{\rm vr}= 0.3$ atm. Since the entire pumping cycle lasts dozens of microseconds, the initial high-pressure phase can be approximated as instantaneous impact on the fluid, as a result of which the fluid acquires mechanical momentum $q_0$ \cite{Asai1992}. This leads to the following impulse bubble model:
\begin{equation}
p_{\rm v}(t) = \frac{q_0}{A} \, \delta(t) + p_{\rm vr} \: .
\label{onedf:eq:seven}
\end{equation}
In this approximation, the bubble strength is characterized by the initial fluid momentum $q_0$. Since the impulse has negligible time duration, $q_0$ drops out from the model equations and enters the dynamics only via the initial conditions. This greatly simplifies the analysis. If a detailed physical model of the vapor bubble is known, $q_0$ can be estimated from Eq.~(\ref{onedf:eq:seven}) by integrating the vapor pressure over the bubble's lifetime. In this work, $q_0$ will be treated as an independent phenomenological parameter.      

Consider now the pressures at the channel's ends $p_{1,2}$. In general, they are different from the respective bulk pressures $p_{1b}$ and $p_{2b}$. The fluid dynamics in the reservoirs is complex. First, the formation of a jet during outward flow and the formation of a sink during inward flows results in asymmetry. Second, the pumping event is inherently transient. The fluid expelled in the reservoirs forms three-dimensional vortices that persist long after the bubble collapse~\cite{Torniainen2012}. One consequence of the vortices is nonhomogeneous inward flow: The fluid is filling up the channel along the walls while emptying the channel near its axis at the same time. All of this makes the task of finding a representative one-dimensional boundary condition nontrivial. However, a detailed analysis of the reservoir flow is beyond the scope of the present work. The main goal here is to elucidate the physics behind inertial pumping by using simplified one-dimensional dynamics as the analysis tool. From this perspective, replacing the complex reservoir dynamics with a deterministic boundary condition at the end of the channel is acceptable, as long as it is physically justified and consistent with net pumping. 

The easiest approach is to simply neglect all the above-mentioned effects and set the end pressures equal to the bulk values: $p_1 = p_{1b}$ and $p_2 = p_{2b}$. This choice will be referred to as the symmetrical model because it does not distinguish between outflow and inflow. The main advantage of the symmetrical model is simplicity; it is the easiest way to derive the main qualitative features of inertial pumping. 

A different model was proposed by Yuan and Prosperetti~\cite{Yuan1999b}, who tried to account for the asymmetry between expansion and collapse. During bubble expansion, the fluid is injected into the reservoir as a jet that separates from the rest of the fluid. The pressure near the entry point is close to the bulk pressure of the reservoir, $p_{i} = p_{ib}$. During collapse, however, part of the pressure difference is expended on accelerating the fluid inside the reservoir. This is a sink flow \cite{Batchelor1967}. According to the Bernoulli equation, $p_{i} = p_{ib} - \frac{1}{2} \rho \dot{x}^2_{i}$. As mentioned above, vortex formation near channel ends disturbs the sink flow and it is unclear to what extent the Bernoulli correction can be applied to the situation at hand. In the absence of a detailed analysis of the pressure distribution, the proposal of Yuan and Prosperetti will be adopted here as the second reasonable choice of the boundary condition. Hereafter, it will be referred to as the asymmetrical model. Compared with the symmetrical model, it provides additional insight into the inertial pumping mechanism, but is more complex mathematically. 

It is of benefit to keep both options available for analysis. To this end, we introduce the discrete index $m = \{ 0, 1 \}$ that will distinguish between the models: $m = 0$ corresponds to the symmetrical model and $m = 1$ to the asymmetrical model. With such notation the pressure boundary conditions can be written as     
\begin{eqnarray}
p_1 & = & p_{1b} - \frac{m}{2} \, \rho \, \dot{x}^2_1 H( \dot{x}_1) \: ,
\label{onedf:eq:eight} \\
p_2 & = & p_{2b} - \frac{m}{2} \, \rho \, \dot{x}^2_2 H(-\dot{x}_2) \: .
\label{onedf:eq:nine}
\end{eqnarray}
Here $H(z)$ is the Heaviside step function: $H(z > 0) = 1$ and $H(z < 0) = 0$.

\subsection{\label{oned:sec:twothree}
Dimensionless pump equations 
}

The number of model parameters in Eqs.~(\ref{onedf:eq:four})--(\ref{onedf:eq:nine}) can be reduced by transforming the equations to a dimensionless form. In a similar procedure in Ref.~[\onlinecite{Torniainen2012}], the channel diameter was taken as the unit of length and the duration of the high-pressure phase as the unit of time. In the present case, both these parameters are zero, so a different pair of units is required. The choice of unit length is obvious: The total channel length $L$ is the only model parameter of that dimensionality. To choose a unit time one can proceed as follows. In the most typical case, the bulk reservoir pressure is equal to the atmospheric pressure $p_0$. The two fluid columns will be evolving under negative pressure difference $-(p_0 - p_{\rm vr})$. That sets a characteristic fluid velocity as $\sqrt{(p_0-p_{\rm vr})/\rho}$ and a characteristic time as $L \sqrt{\rho/(p_0-p_{\rm vr})}$. The characteristic time thus defined is of the order of the total bubble lifetime and the characteristic velocity is approximately the mean fluid velocity over the same period. Accordingly, one introduces dimensionless time $\tau$ and interface positions $\xi_{i}$
\begin{eqnarray}
 \tau     & = & \frac{t}{L} \sqrt{\frac{p_0-p_{\rm vr}}{\rho}} \: ,
\label{onedf:eq:ten} \\
\xi_{1,2} & = & \frac{x_{1,2}}{L} \: .
\label{onedf:eq:eleven}
\end{eqnarray}
Upon substitution of Eqs.~(\ref{onedf:eq:seven})--(\ref{onedf:eq:eleven}), the pump equations (\ref{onedf:eq:four}) and (\ref{onedf:eq:five}) become
\begin{eqnarray}
      \xi_1   \, \xi''_1 + \frac{m}{2} \xi'^2_1 H( \xi'_1) + \beta \,       \xi_1   \, \xi'_1 & = &   \gamma_1 ,
\label{onedf:eq:twelve} \\
( 1 - \xi_2 ) \, \xi''_2 - \frac{m}{2} \xi'^2_2 H(-\xi'_2) + \beta \, ( 1 - \xi_2 ) \, \xi'_2 & = & - \gamma_2 ,
\label{onedf:eq:thirteen}
\end{eqnarray}
where $0 < \xi_1(\tau) \leq \xi_2(\tau) < 1$, the primes denote differentiation with respect to $\tau$, and
\begin{eqnarray}
 \beta       & \equiv & \frac{\kappa L}{\rho A} \sqrt{\frac{\rho}{p_0-p_{\rm vr}}} \: ,
\label{onedf:eq:fourteen} \\
\gamma_{1,2} & \equiv & \frac{p_{1b,2b} - p_{\rm vr}}{p_{0} - p_{\rm vr}}      \: .
\label{onedf:eq:fifteen}
\end{eqnarray}
The equations of motion contain one discrete model parameter $m = \{ 0,1 \}$ and three positive continuous dimensionless parameters: friction $\beta$ and pressures $\gamma_{1}$ and $\gamma_{2}$. In the most common case of bulk pressures equal to the atmospheric pressure $\gamma_1 = \gamma_2 = 1$, only one continuous parameter $\beta$ remains. 

Consider now the initial conditions. The expansion phase starts with a zero-size bubble located at $x_0$, hence 
\begin{equation}
\xi_1(0) = \xi_2(0) = \frac{x_0}{L} = \xi_0 \: .
\label{onedf:eq:sixteen}
\end{equation}
The microheater location $\xi_0$ is a critical parameter of the inertial pumping effect. For a symmetrically placed microheater $\xi_0 = 1/2$, the net flow must be zero. The net flow increases as $\xi_0$ deviates from $1/2$, but in a nonmonotonic manner. This behavior will be analyzed in the following sections. 

Another important parameter is the bubble strength. Within the impulse-bubble approximation, at $t = 0$ the two columns of fluid acquire momentum $q_0$. Converting this to initial velocities, one obtains $v_1(0) = - q_0/\rho A x_0$ and $v_2(0) = q_0/\rho A (L-x_0)$. In dimensionless units       
\begin{equation}
\xi'_1(0) = - \frac{\alpha}{\xi_0} \: , \hspace{0.5cm} \xi'_2(0) = \frac{\alpha}{1 - \xi_0} \: ,
\label{onedf:eq:seventeen}
\end{equation}
where
\begin{equation}
\alpha \equiv \frac{q_0}{\rho A L} \sqrt{\frac{\rho}{p_{0} - p_{\rm vr}}} \: .
\label{onedf:eq:eighteen}
\end{equation}
The initial conditions (\ref{onedf:eq:seventeen}) assume that expansion begins with a state of rest. One may consider a more general case of fluid already moving as a whole at $\tau = 0$. (It occurs, for example, with high-frequency repeated pumping or when pumping against an imposed pressure gradient.) This case is not included in the present work. 

To summarize, bubble kinematics is completely defined by the positions of vapor-fluid interfaces satisfying $0 < \xi_1 \leq \xi_2 < 1$. Bubble dynamics is governed by the equations of motion (\ref{onedf:eq:twelve}) and (\ref{onedf:eq:thirteen}) and initial conditions (\ref{onedf:eq:sixteen}) and (\ref{onedf:eq:seventeen}). There are five continuous dimensionless parameters $\alpha$, $\beta$, $\gamma_{1}$, $\gamma_{2}$, and $\xi_0$ and one discrete model index $m$. Typical parameter values are given in Table~\ref{onedf:tab:one}.

\begin{table}
\renewcommand{\tabcolsep}{0.2cm}
\renewcommand{\arraystretch}{1.5}
\begin{tabular}{|c|c|c|}
\hline\hline
 Parameter       & Meaning             & Typical values     \\ \hline
\hline 
 $\alpha$        & bubble strength     & 0.1--1.0           \\ \hline
 $\beta$         & friction            &   1--10            \\ \hline
 $\gamma_{1,2}$  & pressure difference &   1--2             \\ \hline
 $\xi_0$         & asymmetry           &  (0,1)             \\ \hline 
\hline 
\end{tabular}
\caption{
Physically relevant ranges of the dimensionless model parameters. The estimates have been obtained for $\rho = 10^3$ kg/m$^3$, $L = 200$ $\mu$m, $A = 20^2$ $\mu{\rm m}^2$, $p_0 = 1$ atm, and $p_{\rm vr} = 0.3$ atm~\cite{Torniainen2012}. Using a high-pressure phase duration of 1 $\mu$s and pressure of 8 atm~\cite{Torniainen2012}, the initial impulse is estimated as $q_0/A \approx 0.7$ kg/m s. Then Eq.~(\ref{onedf:eq:eighteen}) yields $\alpha \approx 0.42$, which is a typical value of the bubble strength. For a fluid with bulk viscosity $\eta = 1.3$ mPa s, Eqs.~(\ref{onedf:eq:fourteen}) and (\ref{onedf:eq:six}) yield $\beta \approx 2.0$, which is adopted as a typical value of the friction coefficient.  
} 
\label{onedf:tab:one}
\end{table}

\subsection{\label{oned:sec:twofour}
Postcollapse phase
}

The collapse phase ends with the two fluidic columns colliding at time $t_c$ at point $x_c$ which is different from the starting point of expansion $x_0$. The shift $\triangle x = x_c - x_0$ constitutes the primary pumping effect. In addition, momenta of the two flows are in general different. (This is despite the fact that motion started with equal momenta $q_0$. Reasons for the eventual inequality are analyzed in the following section.) By momentum conservation this implies that after collapse the fluid will continue to flow as a whole (i.e., with total length $L$ and mass $\rho A L$) until it is brought to a complete stop by the pressure difference and viscous forces. The velocity $v_c$ in the beginning of this phase will be referred to as postcollapse velocity and net flow during the postcollapse phase as the secondary pumping effect. Flow kinematics can still be described by time evolution of collapse point $x(t)$. Acting forces no longer include the bubble vapor pressure but only end-channel pressures $p_{1,2}$. The equation of motion reads  
\begin{equation}
\rho A L \, \ddot{x} + \kappa L \, \dot{x} = ( p_1 - p_2 ) A \: ,
\label{onedf:eq:twenty}
\end{equation}
with initial conditions $x(t_c) = x_c$ and $\dot{x}(t_c) = v_c$. Substituting the pressure model (\ref{onedf:eq:eight}) and (\ref{onedf:eq:nine}) and converting to the dimensionless units, one obtains  
\begin{equation}
\xi'' + \frac{m}{2} \, \xi'^2 \, {\rm sgn}(\xi') + \beta \, \xi' = \gamma_1 - \gamma_2 \: ,
\label{onedf:eq:twentyone}
\end{equation}
with $\xi(\tau_c) = \xi_c$ and $\xi'(\tau_c) = \eta_c$. The parameters in Eq.~(\ref{onedf:eq:twentyone}) have the same meaning as during the expansion-collapse cycle.

\section{\label{onedf:sec:three}
Inertial pumping effect
}

Quite generally, inertial pumping happens because the shorter fluidic column with its smaller mass turns around faster than the longer one under the same pressure difference. The shorter column returns to the starting point earlier and has extra time before collision to keep moving in the long arm's direction. This results in the primary effect. For the same reason, the shorter arm has more time to accelerate during collapse. Consequently, it arrives at collision with a larger momentum than the longer arm, leading to the secondary effect. Qualitative features of inertial pumping can be understood by analyzing return times and interface trajectories near the collision point. In this section we study the basic inviscid pumping event with $\beta = 0$ and $\gamma_1 = \gamma_2 = 1$.

\subsection{\label{onedf:sec:threeone}
Expansion phase
}

\begin{figure}[t]
\includegraphics[width=0.48\textwidth]{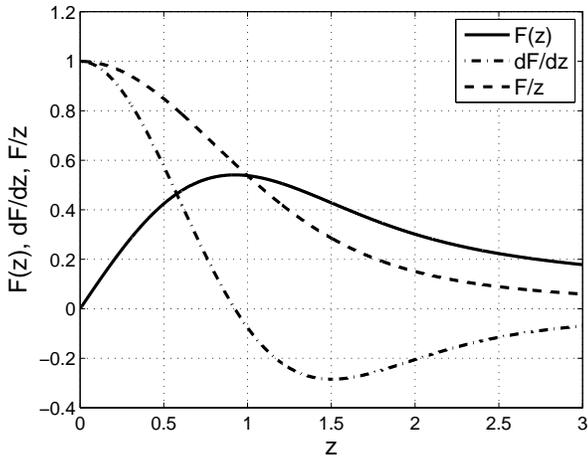}
\caption{Dawson integral (\ref{onedf:eq:twentysix}), its derivative, and the function $\frac{1}{z}F(z)$. Asymptotic expansions are $F(z \ll 1) \approx z - \frac{2}{3} z^3 + \frac{4}{15} z^5 + \cdots$ and $F(z \gg 1) \approx \frac{1}{2z} + \frac{1}{4z^3} + \cdots$.}   
\label{oned:fig:two}
\end{figure}

It is sufficient to follow the dynamics of only one vapor-fluid interface. The math is slightly more transparent for $\xi_1$. The initial value problem reads [cf. Eq.~(\ref{onedf:eq:twelve}) for $\xi'_1 < 0$]
\begin{equation}
\left\{ \begin{array}{l}
\xi_1 \xi''_1             = 1                           \\
\xi_1(0)                  = \xi_0                       \\
\xi'_1(0)                 = - \frac{\alpha}{\xi_0} 
\end{array} \right.                                     \: .                           
\label{onedf:eq:twentytwo}
\end{equation}
Integrating once, one obtains a first integral
\begin{equation}
\frac{1}{2} \, \xi'^2_1 - \ln{\xi_1} = C = 
\frac{1}{2} \left( \frac{\alpha}{\xi_0} \right)^2 - \ln{\xi_0} \: .
\label{onedf:eq:twentythree}     
\end{equation}
The interface undergoes potential motion in a profile $-\ln{\xi}$, with $C$ being analogous to a total energy. At a turning point $\xi'_1 = 0$, from where the coordinate of the turning point is
\begin{equation}
\xi_{1t} = \xi_0 \: e^{- \frac{1}{2} \left( \frac{\alpha}{\xi_0} \right)^2} .
\label{onedf:eq:twentyfour}     
\end{equation}
Integrating a second time from the start to the turning point, the total expansion time is given by
\begin{equation}
\tau_{1e} = \int^{\xi_0}_{\xi_{1t}} \frac{d\xi}{\sqrt{2 \ln{\frac{\xi}{\xi_{1t}}} }} = 
\sqrt{2} \xi_0 \: e^{- \frac{1}{2} \left( \frac{\alpha}{\xi_0} \right)^2} 
\int^{\frac{\alpha}{\sqrt{2} \xi_0}}_0 e^{z^2} dz \: .
\label{onedf:eq:twentyfive}     
\end{equation}
Introducing the Dawson integral (Fig.~\ref{oned:fig:two}) 
\begin{equation}
F(z) = e^{-z^2} \int^{z}_{0} e^{u^2} du  \: ,
\label{onedf:eq:twentysix}
\end{equation}
the expansion time can be written as
\begin{equation}
\tau_{1e} = \alpha \: \frac{1}{\left[ \frac{\alpha}{\sqrt{2} \, \xi_0} \right] } \, 
F \left[ \frac{\alpha}{\sqrt{2} \xi_0} \right] \: .
\label{oned:eq:twentyseven}     
\end{equation}
The expansion time of the right interface is obtained from here by substitution $\xi_0 \rightarrow 1 - \xi_0$: 
\begin{equation}
\tau_{2e} = \alpha \: \frac{1}{\left[ \frac{\alpha}{\sqrt{2} \, ( 1 - \xi_0) } \right] } \, 
F \left[ \frac{\alpha}{\sqrt{2} ( 1 - \xi_0 ) } \right] \: .
\label{onedf:eq:twentyeight}     
\end{equation}
A crucial observation now is that the function $\frac{1}{z} F(z)$ monotonically decreases with its argument, as shown in Fig.~\ref{oned:fig:two}. Clearly, expansion times are the same for the symmetrical microheater position $\xi_0 = 1/2$. However, for asymmetric positions (and the same bubble strength) the expansion times will be different: $\tau_{1e} < \tau_{2e}$ for $\xi_0 < 1/2$ and $\tau_{1e} > \tau_{2e}$ for $\xi_0 > 1/2$. Thus, due to its smaller mass, the shorter arm slows down faster and reaches the turnaround point earlier than the longer one. This forms the basis for inertial effects of both types.

\subsection{\label{onedf:sec:threetwo}
Collapse phase. Symmetrical model ($m = 0$)
}

In the symmetrical model collapse dynamics is an inverse of expansion dynamics. Acceleration can be described as sliding along the same potential profile $-\ln{\xi}$ with a zero starting velocity. The return times (i.e., times after which both interfaces return to the starting point $\xi_0$) are simply twice the expansion times $\tau_{1r,2r} = 2 \tau_{1e,2e}$. Both the primary and secondary pumping effects can be derived analytically for weak and strong asymmetries. These cases are considered first. After that a complete numerical solution of the pump equations is presented.

\subsubsection{\label{onedf:sec:threetwoone}
Weak asymmetry
}

Under weak asymmetry, the microheater is close to the channel's midpoint:
\begin{equation}
\xi_0 = \frac{1}{2} + \psi \: , \hspace{1.0cm} \psi \ll 1  \: .
\label{onedf:eq:twentynine}     
\end{equation}
We are interested in the pumping effects that are of first order in $\psi$. Substituting $\xi_0$ in the return times and expanding to first order, one obtains
\begin{eqnarray}
\tau_{1r,2r} & = & a \pm b \, \psi + O(\psi^2) \: , 
\label{onedf:eq:thirty}           \\
 a           & = & \sqrt{2} F(\sqrt{2} \alpha) \: , 
\label{onedf:eq:thirtyone}           \\
 b           & = & 2^{3/2}  F(\sqrt{2} \alpha) - 4 \alpha F'(\sqrt{2} \alpha)  
                   \nonumber         \\ 
             & = & 2^{3/2} ( 1 + 4 \alpha^2 ) F(\sqrt{2} \alpha) - 4 \alpha   \: .
\label{onedf:eq:thirtytwo}     
\end{eqnarray}
The last line follows from the identity $F'(z) = 1 - 2 z F(z)$. The limit values for weak and strong bubbles are
\begin{eqnarray}
 a (\alpha \ll 1 ) & \approx & 2 \alpha - \frac{8\alpha^3}{3} \: , \hspace{0.5cm} 
 a (\alpha \gg 1)    \approx   \frac{1}{2\alpha} \: ,
\label{oned:eq:seventy}           \\
 b (\alpha \ll 1 ) & \approx & \frac{32\alpha^3}{3}           \: , \hspace{1.2cm} 
 b (\alpha \gg 1 )   \approx   \frac{2}{\alpha}  \: ,
\label{onedf:eq:thirtythree}     
\end{eqnarray}
from where asymptotes of the return times can be deduced. Of primary interest are the time and coordinate of the collision point. They can be found by linearizing interface trajectories near the starting point just prior to collision. At $\tau = \tau_{1r}$, the left interface is at $\xi_1 = \xi_0$ and has velocity $\xi'_1 = \alpha/(\frac{1}{2} + \psi)$. Accordingly, a linearized trajectory is described by the equation
\begin{equation}
\xi_1(\tau) = \left( \frac{1}{2} + \psi \right) + \frac{\alpha}{\frac{1}{2}+\psi} 
\left( \tau - a - b \, \psi \right)  \: .
\label{onedf:eq:thirtyfour}     
\end{equation}
Similarly, the right interface's trajectory linearized aro\-und the collision time is 
\begin{equation}
\xi_2(\tau) = \left( \frac{1}{2} + \psi \right) - \frac{\alpha}{\frac{1}{2}-\psi} 
\left( \tau - a + b \, \psi \right)  \: .
\label{onedf:eq:thirtyfive}     
\end{equation}
Equating $\xi_1(\tau_c) = \xi_2(\tau_c)$ yields the collision time and position in the first order in $\psi$:
\begin{eqnarray}
\tau_c  & = & a + 0 \cdot \psi + O(\psi^2) \: ,
\label{onedf:eq:thirtysix}     \\
\xi_c   & = & \left( \frac{1}{2} + \psi \right) - 2 \alpha b \, \psi \: .
\label{onedf:eq:thirtyseven}     
\end{eqnarray}
The primary pumping effect is the displacement of the collapse point relative to the starting point of expansion. According to Eq.~(\ref{onedf:eq:thirtyseven}), the primary effect is given by
\begin{equation}
\Delta \xi = \xi_c - \xi_0 = - 2 \alpha b \, \psi  = \left\{ 
\begin{array}{ll}
- \frac{64}{3} \, \alpha^4 \psi \: , & \alpha \ll 1 \:   \\
                    - 4 \, \psi \: , & \alpha \gg 1 \: . 
\end{array} \right. 
\label{onedf:eq:thirtyeight}     
\end{equation}
Note that the primary effect is a sharp function of the bubble strength (proportional to $\alpha^4$) for weak bubbles. 

The secondary pumping effect originates from the imbalance of mechanical momenta at collision. The velocities can also be found from linearizing them around the collision point. Accelerations follow from the original dynamic equations. The linearized velocities are
\begin{eqnarray}
\xi'_{1c} & = & \frac{\alpha}{\frac{1}{2}+\psi} + 
\frac{1}{\frac{1}{2}+\psi} ( \tau_c - \tau_{1r} ) = \frac{\alpha - b \, \psi}{\frac{1}{2}+\psi} \: ,
\label{onedf:eq:thirtynine}     \\
\xi'_{2c} & = & \frac{-\alpha}{\frac{1}{2}-\psi} - 
\frac{1}{\frac{1}{2}-\psi} ( \tau_c - \tau_{2r} ) = \frac{- \alpha - b \, \psi}{\frac{1}{2}-\psi} \: .
\label{onedf:eq:forty}     
\end{eqnarray}

The total momentum or velocity after collapse to the first order in $\psi$ is
\begin{eqnarray}
\eta_c & = & \xi_c \: \xi'_{1c} + ( 1 - \xi_c ) \: \xi'_{2c} 
         = - 2 b ( 1 + 4 \alpha^2 ) \psi          \nonumber \\
       & = & \left\{ \begin{array}{ll} 
- \frac{64}{3} \, \alpha^3 \psi \: , & \alpha \ll 1 \: , \\    
            - 16 \, \alpha \psi \: , & \alpha \gg 1 \: .    
\end{array} \right.  
\label{oned:eq:eighty}     
\end{eqnarray}
Since the total mass after collapse is equal to 1, the postcollapse momentum and velocity are given by the same expression. The first term in Eq.~(\ref{oned:eq:eighty}) originates from the additional velocity that the short arm acquires between the return time and collision time. The second term comes from an additional mass that the short arm acquires during the same time interval. Depending on bubble strength $\alpha$, either the first or the second factor dominates the secondary effect.

\subsubsection{\label{onedf:sec:threetwotwo}
Strong asymmetry
}

Consider now the opposite case of strong asymmetry when the microheater is very close to one edge of the channel, say, to the left one. Then the starting point $\xi_0 \ll 1$ is a small parameter. Of interest are the primary and secondary pumping effects in the lowest order of $\xi_0$.  

Let us start with the return times. Clearly, it takes the short arm a much smaller time to return than the long arm. Referring to Eq.~(\ref{oned:eq:twentyseven}) and using the large-argument expansion $F(z \gg 1) = (2z)^{-1} + O(z^{-3})$, one obtains
\begin{equation}
\tau_{1r}(\xi_0 \ll 1) = 2 \tau_{1e}(\xi_0 \ll 1) = \frac{2 \xi^2_0}{\alpha} \: .
\label{oned:eq:eightyone}     
\end{equation}
Thus the short arm's return time is second order in $\xi_0$. One power comes from a short travel distance and another power from a large starting velocity. In contrast, the return time of the long arm is $O(1)$.   

\begin{figure}[t]
\includegraphics[width=0.48\textwidth]{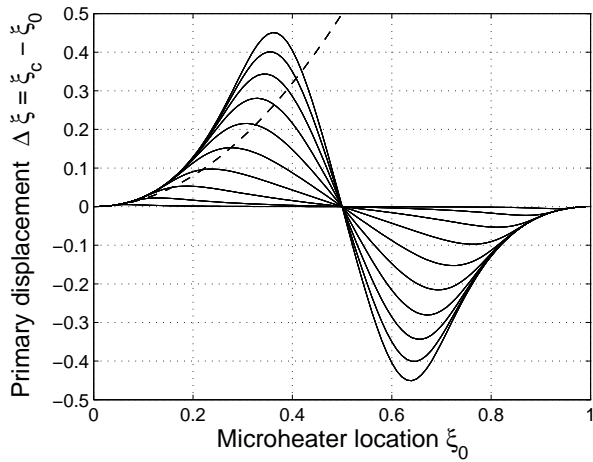}
\caption{Primary displacement of the collapse point $\Delta \xi = \xi_c - \xi_0$ as a function the starting point $\xi_0$ (microheater location) as determined from numerical solution of the dimensionless pump equations (\ref{onedf:eq:twelve}) and (\ref{onedf:eq:thirteen}) for $\beta = m = 0$ and $\gamma_1 = \gamma_2 = 1$. Different lines correspond to different bubble strengths $\alpha = 0.1, 0.2, \ldots , 1.0$. The dashed line is the strong-asymmetry asymptote (\ref{oned:eq:eightysix}).}   
\label{oned:fig:three}
\end{figure}

The collision time and position can again be found from linearized trajectories around the collision point. At $\tau = \tau_{1r}$, the short arm has velocity $\alpha/\xi_0$ and acceleration $1/\xi_0$. The trajectory is 
\begin{equation}
\xi_1(\tau) = \xi_0 + \frac{\alpha}{\xi_0} ( \tau - \tau_{1r} ) + \frac{1}{2\xi_0} ( \tau - \tau_{1r} )^2 + \cdots .
\label{oned:eq:eightytwo}     
\end{equation}
The right arm moves comparatively slow; therefore, it is sufficient to expand its trajectory around the starting point and time. It reads:
\begin{equation}
\xi_2(\tau) = \xi_0 + \frac{\alpha}{1 - \xi_0} \, \tau - \frac{1}{2 ( 1 - \xi_0 ) } \, \tau^2 + \cdots .
\label{oned:eq:eightythree}     
\end{equation}
By setting $\xi_1(\tau_c) = \xi_2(\tau_c)$ after some algebra one obtains the time and position of the collision point
\begin{eqnarray}
\tau_c  & = & \tau_{1r} ( 1 + \xi_0 ) + \cdots = \frac{2 \xi^2_0}{\alpha} ( 1 + \xi_0 ) + O(\xi^4_0) \: ,
\label{oned:eq:eightyfour}     \\
\xi_c   & = & \xi_0 + 2 \xi^2_0 + \cdots . 
\label{oned:eq:eightyfive}     
\end{eqnarray}
Thus the primary pumping effect is 
\begin{equation}
\Delta \xi = \xi_c - \xi_0 = 2 \xi^2_0 \: ,  
\label{oned:eq:eightysix}     
\end{equation}
which is quadratic in $\xi_0$. It is shown in Fig.~\ref{oned:fig:three} by the dashed line. 

The velocities needed for the secondary effect can be found from the same trajectories (\ref{oned:eq:eightytwo}) and (\ref{oned:eq:eightythree}):
\begin{eqnarray}
\xi'_{1c}  & = & \frac{\alpha}{\xi_0} + \frac{1}{\xi_0} (\tau_c - \tau_{1r}) = 
\frac{\alpha}{\xi_0} + \frac{2 \xi^2_0}{\alpha} \: ,
\label{oned:eq:eightyseven}     \\
\xi'_{2c}  & = & \frac{\alpha}{1 - \xi_0} + \frac{\tau_c}{1 - \xi_0} =  
\frac{\alpha - \frac{2\xi^2_0}{\alpha}(1+\xi_0) }{1 - \xi_0} \: . 
\label{oned:eq:eightyeight}     
\end{eqnarray}
The postcollapse velocity is
\begin{equation}
\eta_c = \xi_c \: \xi'_{1c} + ( 1 - \xi_c ) \: \xi'_{2c} = 2 \alpha ( 1 + \xi_0 ) + O(\xi^2_0) \: .
\label{oned:eq:eightynine}     
\end{equation}
The linear $\xi_0$ term comes from the increased mass that the short arm picks up between return time $\tau_{1r}$ and collision time $\tau_c$.

\begin{figure}[t]
\includegraphics[width=0.48\textwidth]{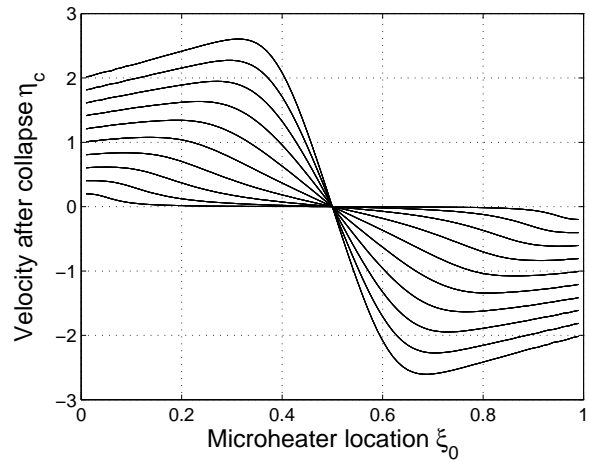}
\caption{Dimensionless postcollapse velocity $\eta_c$ as determined from numerical solution of the dimensionless pump equations (\ref{onedf:eq:twelve}) and (\ref{onedf:eq:thirteen}) for $\beta = m = 0$ and $\gamma_1 = \gamma_2 = 1$. The lines correspond to different bubble strengths $\alpha = 0.1, 0.2, \ldots , 1.0$.}   
\label{oned:fig:four}
\end{figure}

Notice that the postcollapse momentum can exceed twice the initial fluid momentum $\alpha$, as can also be seen from the exact numerical solution presented in Fig.~\ref{oned:fig:four}. The reason for this excess of momentum is as follows. In the symmetrical model, the collapse dynamics is simply the reversal of expansion. The short arm returns to the starting position with a momentum equal in magnitude but opposite in sign to its initial value. If $\xi_0 \ll 1$, the return time is very small, so the long arm has lost very little of its initial momentum. By the time the short arm returns to $\xi_0$, the combined momentum is almost (slightly less than) $2\alpha$. However, since the long arm has moved away from $\xi_0$, the short arm has a little more time to accelerate before collision. During this extra time the short arm picks up more momentum than the long one loses, which results in an excess. Ultimately, the momentum is provided by the left reservoir.

\subsubsection{\label{onedf:sec:threetwothree}
Complete numerical solution
}

\begin{figure}[t]
\includegraphics[width=0.48\textwidth]{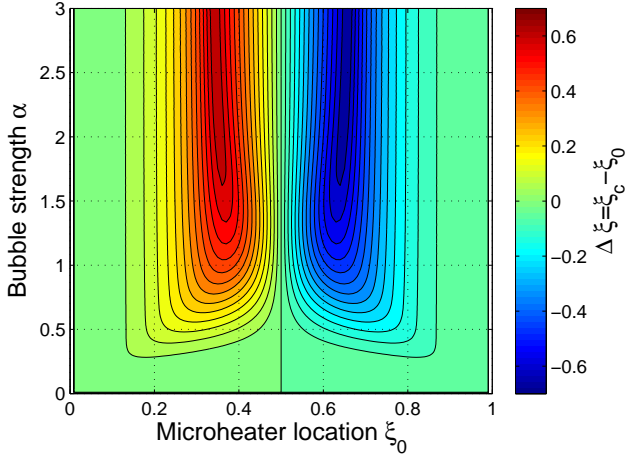}
\caption{Complete solution of the pump equations (\ref{onedf:eq:twelve}) and (\ref{onedf:eq:thirteen}) for $\beta = m = 0$ and $\gamma_1 = \gamma_2 = 1$. Shown is the primary pumping effect $\Delta \xi = \xi_c - \xi_0$ as a function of microheater location $\xi_0$ and bubble strength $\alpha$. The step between contour lines is $\Delta \xi = 0.05$.}   
\label{oned:fig:six}
\end{figure}

\begin{figure}[t]
\includegraphics[width=0.48\textwidth]{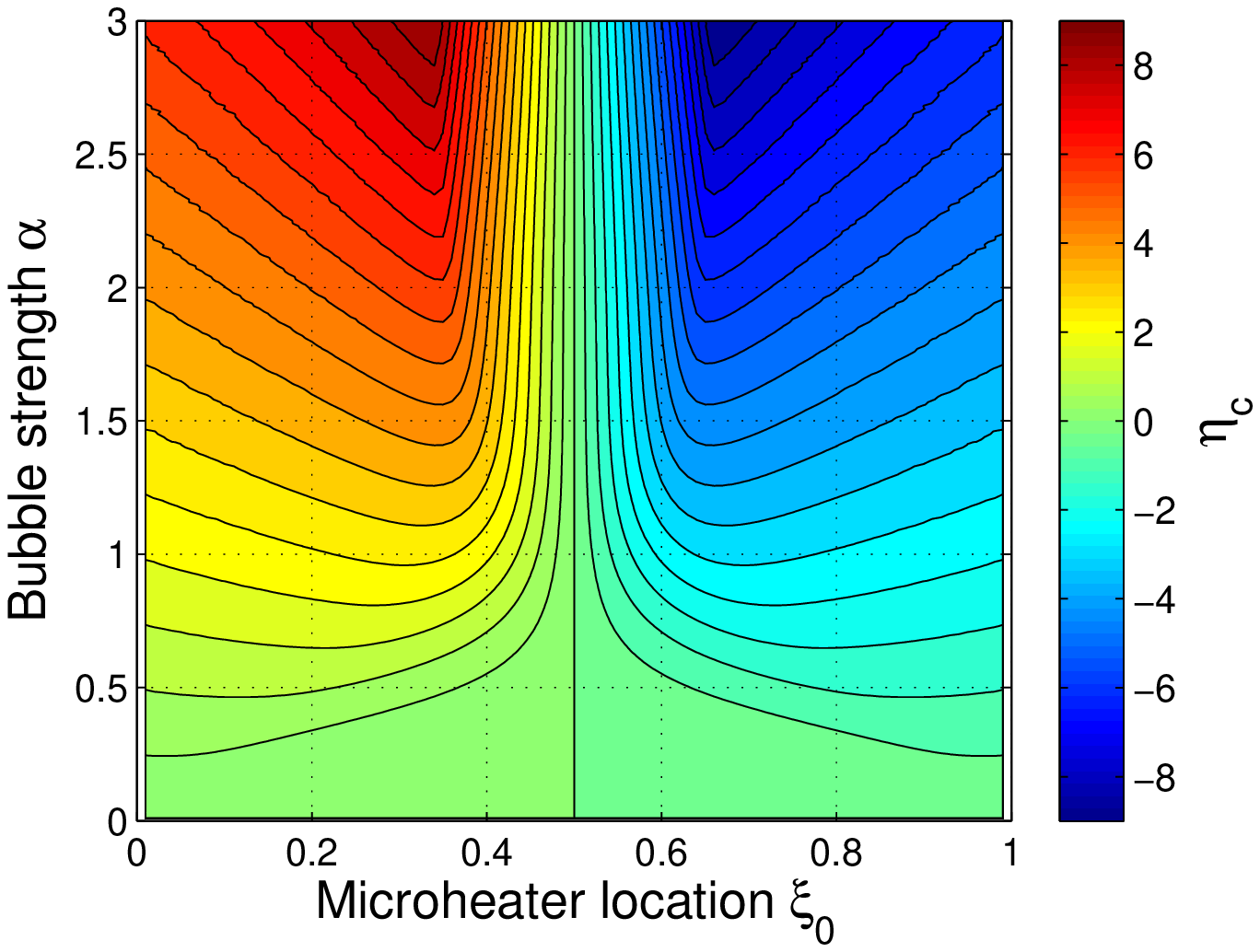}
\caption{Complete solution of the pump equations (\ref{onedf:eq:twelve}) and (\ref{onedf:eq:thirteen}) for $\beta = m = 0$ and $\gamma_1 = \gamma_2 = 1$. Shown is the secondary pumping effect: postcollapse velocity $\eta_c$ as a function of microheater location $\xi_0$ and bubble strength $\alpha$. The step between contour lines is $\Delta \eta = 0.5$.}   
\label{oned:fig:seven}
\end{figure}

With just two dimensionless parameters, the system can be easily analyzed numerically to completion. The numerical solution is presented in Figs.~\ref{oned:fig:three}--\ref{oned:fig:seven}. Figures~\ref{oned:fig:three} and \ref{oned:fig:four} show the primary and secondary effects for a number of bubble-strength values spanning the practically relevant interval $0.1 < \alpha < 1$. The most notable feature of both sets of graphs is an optimal microheater location at which the effects are maximal. Note that, although the optimal $\xi_0$ for the two effects are not exactly equal, they are close and share a similar trend: The optimal location is close to the channel's edge for weak bubbles and shifts toward the channel center as bubbles grow stronger. The existence of an optimal microheater location is a welcome feature for practical applications of the effect. Clearly, fabricating microheaters right at the channel's edge would be challenging. The findings imply that this is unnecessary.      

Figures \ref{oned:fig:six} and \ref{oned:fig:seven} show both inertial effects for the entire parameter space $0 < \xi_0 < 1$ and $0.01 \leq \alpha \leq 3.0$ as color maps. Bubble strengths $\alpha > 1$ are considered impractical, but are still of interest from the fundamental standpoint and are included here as such. Several features are worth noting. (i) For both effects, the optimal microheater location converges to $\xi_0 \approx 0.35, 0.65$ at $\alpha > 1$. (ii) The primary effect saturates with bubble strength, although the $\alpha$ dependence is not monotonic. (iii) The secondary effect does not saturate with bubble strength. It grows roughly linearly at large $\alpha$, in agreement with Eqs.~(\ref{oned:eq:eighty}) and (\ref{oned:eq:eightynine}). We have also verified that the numerical solution agrees with the analytical asymptotes derived earlier in this section.

\subsection{\label{onedf:sec:threethree}
Collapse phase. Asymmetrical model ($m = 1$)
}

In the asymmetrical model, collapse is not simply a time reversal of expansion. Consider the left fluid-vapor interface. After time $\tau_{1e}$ [Eq.~(\ref{onedf:eq:twentyfive})] flow stops at $\xi = \xi_{1t}$ given by Eq.~(\ref{onedf:eq:twentyfour}). After that, collapse proceeds according to the equation    
\begin{equation}
\xi_1 \xi''_1 + \frac{1}{2} \xi'^2_1 = 1 \: ,
\label{onedf:eq:ninety}     
\end{equation}
with a zero initial velocity. The Bernoulli correction results in a portion of pressure difference being spent on accelerating fluid within the reservoir. The implications are more pronounced whenever high collapse velocities are encountered, i.e., for strong asymmetries and strong bubbles. Indeed, when the mass at the turning point is small it accelerates quickly to a velocity of $\sim 1$. At this moment the Bernoulli correction kicks in, the pressure difference within the channel drops and acceleration slows down. The velocity approaches the limit value of $\sqrt{2}$. New features of the asymmetrical model discussed below are consequences of this fact.         

In the equation of motion (\ref{onedf:eq:ninety}) both integrations are elementary with the results
\begin{equation}
\frac{1}{2} \xi'^2_1 + \frac{\xi_{1t}}{\xi_1} = 1 \: ,
\label{onedf:eq:ninetyone}     
\end{equation}
\begin{equation}
\sqrt{\xi_1(\xi_1-\xi_{1t})} + 
\xi_{1t} \ln{\frac{\sqrt{\xi_1} + \sqrt{\xi_1 - \xi_{1t}}}{\sqrt{\xi_{1t}}}} 
= \sqrt{2} (\tau - \tau_{1e}) \: .
\label{onedf:eq:ninetytwo}     
\end{equation}
The left edge's return velocity and time are obtained from here by setting $\xi_1 = \xi_0$:
\begin{equation}
\xi'_{1r} = \sqrt{2 (1 - \epsilon) } \: .
\label{onedf:eq:ninetythreeone}     
\end{equation}
\begin{equation}
\tau_{1r} = \tau_{1e} + \frac{\xi_0}{\sqrt{2}} \left( 
\sqrt{1 - \epsilon} + \epsilon \ln{\frac{1 + \sqrt{1 - \epsilon}}{\sqrt{\epsilon}}}
\right) \: ,
\label{onedf:eq:ninetythree}     
\end{equation}
where $\epsilon = \exp{ ( - \alpha^2 /2 \xi^2_0 ) }$. Notice that the return velocity $\xi'_{1r}$ is equal to the initial velocity $\alpha/\xi_0$ only in the case of weak bubbles $\alpha \ll 1$. 

The weak-asymmetry and strong-asymmetry limit cases are discussed next.

\subsubsection{\label{onedf:sec:threethreeone}
Weak asymmetry
}

Using $\xi_0 = \frac{1}{2} + \psi$ [Eq.~(\ref{onedf:eq:twentynine})] and expanding for $\psi \ll 1$, one obtains after some algebra the return times
\begin{equation}
\tau_{1r,2r} = c(\alpha) \pm g(\alpha) \, \psi + O(\psi^2) \: , 
\label{onedf:eq:ninetyfour}            
\end{equation}
\begin{eqnarray}
 c   & = & \frac{a}{2} + 
 \frac{ \sqrt{1 - \epsilon_0} + \epsilon_0 \ln{(1 + \sqrt{1 - \epsilon_0})} 
       + \alpha^2 \epsilon_0 }{2^{3/2}} \nonumber \\ 
     & = &  \left\{ \begin{array}{ll}
   2 \alpha - \frac{5}{2} \alpha^3 + \cdots        \: , & \alpha \ll 1     \\
   \frac{1}{2^{3/2}} + \frac{1}{4 \alpha} + \cdots \: , & \alpha \gg 1 \: ,  
  \end{array}   \right.         
\label{onedf:eq:ninetyfive}            
\end{eqnarray}
\begin{eqnarray}
 g  & = & \frac{b}{2} + \frac{\sqrt{1 - \epsilon_0} 
          + \epsilon_0 \ln{(1 + \sqrt{1 - \epsilon_0})}}{\sqrt{2}} 
                                                        \nonumber           \\ 
    &   & + \sqrt{2} \epsilon_0 \alpha^2 \left\{ 2\alpha^2 - \frac{1}{2} + 
            2 \ln{(1 + \sqrt{1 - \epsilon_0})} \right.    \nonumber         \\
    &   & \left. - \frac{1}{\sqrt{1-\epsilon_0}} 
- \frac{\epsilon_0}{(1 + \sqrt{1-\epsilon_0}) \sqrt{1-\epsilon_0}} \right\} 
                                                        \nonumber           \\
    & = & \left\{ \begin{array}{ll}
   10 \, \alpha^3                        + \cdots \: , & \alpha \ll 1       \\
   \frac{1}{\sqrt{2}} + \frac{1}{\alpha} + \cdots \: , & \alpha \gg 1  \: , 
  \end{array}   \right.                                                     
\label{onedf:eq:ninetysix}     
\end{eqnarray}
where $\epsilon_0 = \exp{(-2 \alpha^2)}$. The return velocities follow from Eq.~(\ref{onedf:eq:ninetythreeone}) for small $\psi$:
\begin{eqnarray}
\xi'_{1r,2r}  & = & \pm \, h_0 - h_1 \psi       \: ,
\label{onedf:eq:ninetyseven}                                                \\ 
  h_0(\alpha) & = &  \sqrt{2 (1 - \epsilon_0) }              \: ,
\label{onedf:eq:ninetyeight}                                                \\
  h_1(\alpha) & = & \frac{4 \sqrt{2} \alpha^2 \epsilon_0}{\sqrt{1-\epsilon_0}}  \: .      
\label{onedf:eq:ninetynine}            
\end{eqnarray}
With return times and velocities at hand, the collision point can be determined by linearizing trajectories around $\xi_0$. Applying the same method as in the symmetrical model, the collision time and coordinate are found to be  
\begin{eqnarray}
\tau_c  & = & c + 0 \cdot \psi + O(\psi^2) \: ,
\label{onedf:eq:onehundred}     \\
\xi_c   & = & \left( \frac{1}{2} + \psi \right) - h_0 \, c \, \psi \: .
\label{onedf:eq:onehone}     
\end{eqnarray}
Thus the primary pumping effect is
\begin{equation}
\Delta \xi = - \sqrt{2 ( 1-\epsilon_0 )} \, g \, \psi  = \left\{ 
\begin{array}{ll}
- 20 \, \alpha^4 \psi \: , & \alpha \ll 1      \\
          - 1 \, \psi \: , & \alpha \gg 1 \: . 
\end{array} \right. 
\label{onedf:eq:onehtwo}     
\end{equation}
Notice that the transition from the weak-bubble to the strong-bubble regime is not monotonic. The coefficient by $\psi$ passes through a minimum approximately equal to $-3.04$ at $\alpha = 1.12$ before converging to its large-$\alpha$ limit of $-1$.

\begin{figure}[t]
\includegraphics[width=0.48\textwidth]{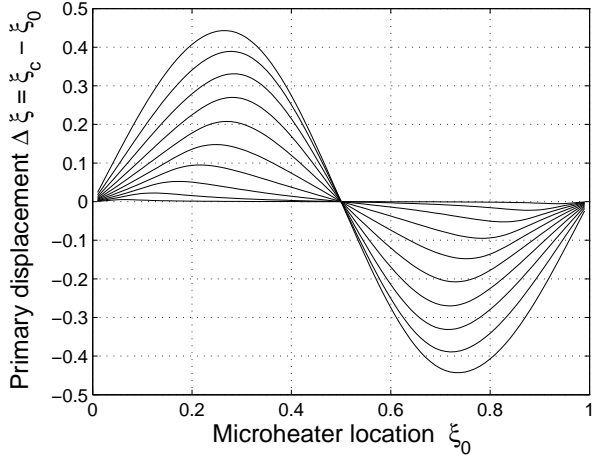}
\caption{Primary displacement $\Delta \xi$ in the asymmetrical pump model (\ref{onedf:eq:twelve}) and (\ref{onedf:eq:thirteen}) for $\beta = 0$ and $\gamma_1 = \gamma_2 = m = 1$. The lines correspond to different bubble strengths $\alpha = 0.1, 0.2, \ldots , 1.0$. }   
\label{oned:fig:eight}
\end{figure}

The postcollapse velocity can be derived from the combined mechanical momentum of the two arms at the time of collision. With the return times and velocities given above and accelerations of both arms known from the original dynamic equations, the velocities near the return times can be approximated by linear functions of time:   
\begin{eqnarray}
\xi'_{1}(\tau) & = &   h_0 - h_1 \psi + \frac{1 - \frac{1}{2} \xi'^2_{1r}}{\xi_0} \, 
  ( \tau - c - g \psi ) \: , 
\label{onedf:eq:onehthree}   \\
\xi'_{2}(\tau) & = & - h_0 - h_ 1\psi + \frac{- 1 + \frac{1}{2} \xi'^2_{2r}}{1 - \xi_0} \, 
  ( \tau - c + g \psi ) \: .   
\label{onedf:eq:onehfour}  
\end{eqnarray}
Substituting here the collision time $\tau = c$, multiplying the velocities by respective column lengths $\xi_c$ and $(1 - \xi_c)$, and adding together, one obtains the total postcollapse momentum or velocity in the leading order in $\psi$:
\begin{equation}
\eta_c = q_{1c} + q_{2c} = \left( 2 h_0 - h^2_0 g - h_1  - 2g \right) \psi \: . 
\label{onedf:eq:onehfive}     
\end{equation}
Here $g$, $h_0$, and $h_1$ are given by explicit expressions in Eqs.~(\ref{onedf:eq:ninetysix}), (\ref{onedf:eq:ninetyeight}), and (\ref{onedf:eq:ninetynine}). The weak-bubble and strong-bubble limits of secondary pumping effect are   
\begin{equation}
\eta_c = \left\{ 
\begin{array}{ll}
 - 16 \, \alpha^3   \, \psi \: , & \alpha \ll 1  \\
 - \frac{4}{\alpha} \, \psi \: , & \alpha \gg 1  \: .
\end{array} \right. 
\label{onedf:eq:onehsix}     
\end{equation}
Notice that unlike the symmetrical model [compare with Eq.~(\ref{oned:eq:eighty})], here there is an optimal bubble strength ($\alpha = 1.05$) for any small microheater asymmetry $\psi$.

\begin{figure}[t]
\includegraphics[width=0.48\textwidth]{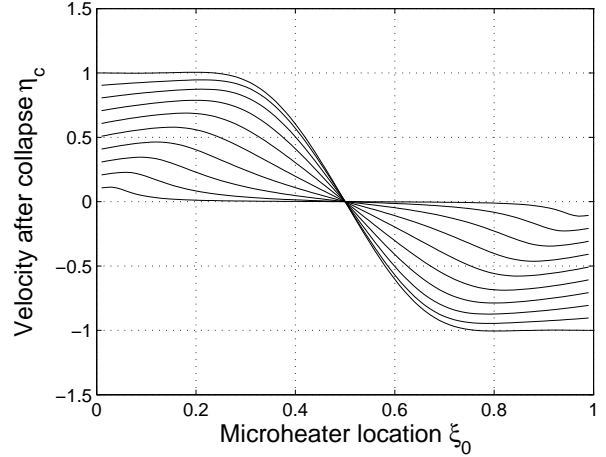}
\caption{Dimensionless postcollapse velocity $\eta_c$ in the asymmetrical pump model (\ref{onedf:eq:twelve}) and (\ref{onedf:eq:thirteen}) for $\beta = 0$ and $\gamma_1 = \gamma_2 = m = 1$. The lines correspond to different bubble strengths $\alpha = 0.1, 0.2, \ldots , 1.0$.}   
\label{oned:fig:nine}
\end{figure}

\subsubsection{\label{onedf:sec:threethreetwo}
Strong asymmetry
}

\begin{figure}[t]
\includegraphics[width=0.48\textwidth]{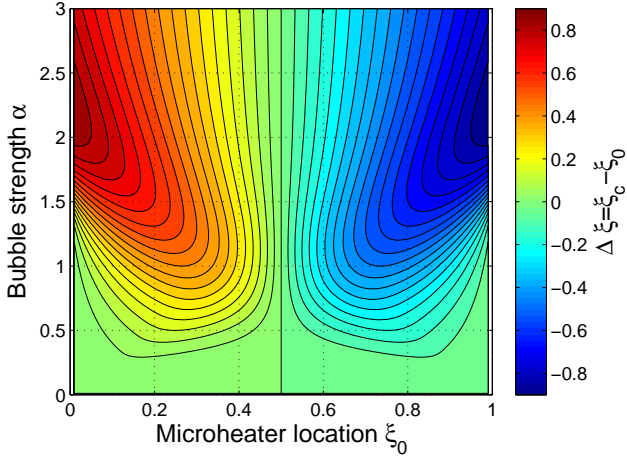}
\caption{Complete solution of the pump equations (\ref{onedf:eq:twelve}) and (\ref{onedf:eq:thirteen}) for $\beta = 0$, $m = 1$, and $\gamma_1 = \gamma_2 = 1$. Shown is the primary pumping effect $\Delta \xi = \xi_c - \xi_0$ as a function of microheater location $\xi_0$ and bubble strength $\alpha$. The step between contour lines is $\Delta \xi = 0.05$.}   
\label{oned:fig:ten}
\end{figure}

Under strong asymmetry, the microheater is close to a channel end $\xi_0 \ll 1$. In this section the condition $\alpha/\xi_0 \gg 1$ will also be assumed, meaning that bubbles cannot be arbitrarily weak. The physics in this regime is dominated by the fact that the short left arm accelerates fast during collapse and quickly reaches the limit velocity of $\sqrt{2}$. After that motion is uniform. Referring to the general expressions (\ref{onedf:eq:ninetythreeone}) and (\ref{onedf:eq:ninetythree}) and taking into account that $\epsilon$ is exponentially small, one obtains the return velocity and time
\begin{eqnarray}
\xi'_{1}  & = &  \sqrt{2}  \: , 
\label{onedf:eq:onehseven}   \\
\tau_{1r} & = & \frac{\xi_0}{\sqrt{2}} + \tau_{1e}   
            =   \frac{\xi_0}{\sqrt{2}} + \frac{\xi^2_0}{\alpha} + \cdots     
\label{onedf:eq:oneheight}  
\end{eqnarray}
Thus the return time is linear in $\xi_0$ rather than quadratic as in the symmetrical model [compare with Eq.~(\ref{oned:eq:eightyone})]. Linearized trajectories near $\tau_{1r}$ are 
\begin{eqnarray}
\xi_{1}(\tau) & = & \xi_0 + \sqrt{2} ( \tau - \tau_{1r} ) \: , 
\label{onedf:eq:onehnine}   \\
\xi_{2}(\tau) & = & \xi_0 + \frac{\alpha}{1-\xi_0} \, \tau \: .   
\label{onedf:eq:onehten}  
\end{eqnarray}
For the purposes of this section, it is not necessary to include the quadratic term in the right edge's trajectory. Equating $\xi_1 = \xi_2$ yields the time and position of the collision point
\begin{eqnarray}
\tau_{c} & = & \frac{\xi_0}{\sqrt{2} - \alpha}             \: , 
\label{onedf:eq:oneheleven}   \\
\xi_{c}  & = & \frac{\sqrt{2} \, \xi_0}{\sqrt{2} - \alpha} \: .   
\label{onedf:eq:onehtwelve}  
\end{eqnarray}
These formulas assume $\alpha < \sqrt{2}$. The singularity is physical. If the initial velocity of the right edge $\alpha/(1-\xi_0) \approx \alpha$ is close to the limit velocity of the left edge $\sqrt{2}$, it takes a long time for the left edge to catch up. Under this condition, the collision time and displacement are no longer small and higher-order terms must be taken into account. Equation~(\ref{onedf:eq:onehtwelve}) yields the primary pumping effect
\begin{equation}
\Delta \xi = \xi_c - \xi_0 = \frac{\alpha \, \xi_0}{\sqrt{2} - \alpha} \: , 
\label{oned:eq:onehthirteen}     
\end{equation}
which is linear in the small parameter $\xi_0$.

\begin{figure}[t]
\includegraphics[width=0.48\textwidth]{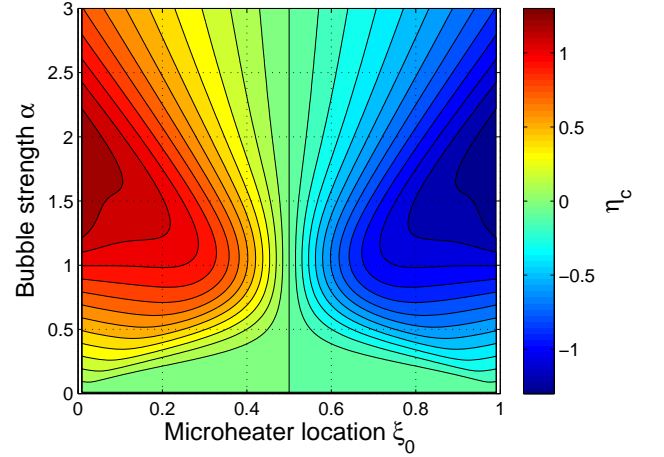}
\caption{Complete solution of the pump equations (\ref{onedf:eq:twelve}) and (\ref{onedf:eq:thirteen}) for $\beta = 0$, $m = 1$, and $\gamma_1 = \gamma_2 = 1$. Shown is the secondary pumping effect: postcollapse velocity $\eta_c$ as a function of microheater location $\xi_0$ and bubble strength $\alpha$. The step between contour lines is $\Delta \eta = 0.1$.}   
\label{oned:fig:eleven}
\end{figure}

The secondary pumping effect derives from the total mechanical momentum at collision
\begin{equation}
\eta_c = \xi_c \sqrt{2} + ( 1 - \xi_c ) \frac{\alpha}{1-\xi_0} = \alpha + \xi_0 (\sqrt{2} + \alpha) \: . 
\label{oned:eq:onehfourteen}     
\end{equation}
The limit value at $\xi_0 \rightarrow 0$ is $\alpha$ and not $2\alpha$ as in the symmetrical model [cf. Eq.~(\ref{oned:eq:eightynine})]. This is because the left arm has a limited velocity. The first-order correction in $\xi_0$ is still positive, which suggests the existence of an optimal microheater location. This conclusion is valid only for not very strong bubbles $\alpha < \sqrt{2}$.

\subsubsection{\label{onedf:sec:threethreethree}
Complete numerical solution
}

A full numerical solution of the asymmetric pump equations is presented in Figs.~\ref{oned:fig:eight}--\ref{oned:fig:eleven}. Referring to Figs.~\ref{oned:fig:eight} and \ref{oned:fig:nine}, several differences from the symmetrical model are apparent. The primary effect is linear at small $\xi_0$ rather than quadratic. At the same time, optimal microheater locations and primary maxima are roughly the same as in the symmetrical model. In contrast, the secondary effect (Fig.~\ref{oned:fig:nine}) is roughly half of the corresponding symmetrical model values. Peaks are less pronounced, but evolve with bubble strength in the same manner. Thus the asymmetrical model predicts smaller net flows. 

Examination of the all-parameter solution of Figs.~\ref{oned:fig:ten} and \ref{oned:fig:eleven} reveals further differences at large bubble strengths. Specifically, the optimal microheater location shifts back toward the channel edge. In addition, both effects feature maxima as functions of $\alpha$, implying an optimal bubble strength for a given microheater location. Finally, the secondary effect does not grow indefinitely at large $\alpha$, unlike in the symmetrical model, but saturates instead. One should note that although these qualitative differences between the two models are intriguing and deserve additional scrutiny, they occur in the regime of very strong vapor bubbles, which is hard to realize in practical devices.       

Summarizing Sec.~\ref{onedf:sec:three}, one concludes that both models studied describe robust inertial pumping. The relative simplicity of the models has allowed many results to be derived analytically, which has elucidated the physics behind the effect. There are enough differences in predictions that a comparison with CFD calculations is likely to  suggest the most realistic model and the underlying channel-reservoir boundary condition. Insights gained from this analysis will be helpful in designing practical inertial pumps.

\begin{figure}[t]
\includegraphics[width=0.48\textwidth]{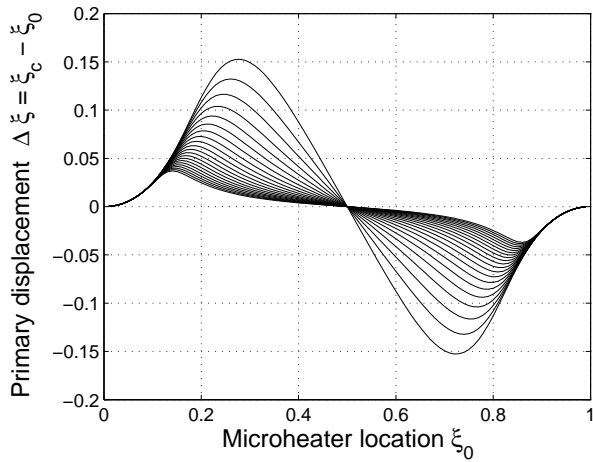}
\caption{Dimensionless primary displacement $\Delta \xi = \xi_c - \xi_0$ in the symmetrical model $m = 0$ for $\gamma_1 = \gamma_2 = 1$ and different friction parameters $\beta = 0.0,0.5,1.0,\ldots,10.0$. The bubble strength is $\alpha = 0.5$. The trace with the largest amplitude corresponds to $\beta = 0.0$. The curves are obtained by numerical solution of the pump equations (\ref{onedf:eq:twelve}) and (\ref{onedf:eq:thirteen}). }   
\label{oned:fig:twelve}
\end{figure}

\section{\label{onedf:sec:four}
Effects of viscosity
}

So far, viscous forces have been neglected. It is intuitively clear that for a mechanism that relies on fluid inertia, viscosity should have a detrimental effect. Within the one-dimensional dynamical model the viscosity is included via the dimensionless parameter $\beta$ defined in Eq.~(\ref{onedf:eq:fourteen}). Here $\beta$ is directly proportional to the bulk viscosity of the participating fluid [cf. Eq.~(\ref{onedf:eq:six})]. To get a sense of the physically relevant $\beta$ interval, a fluid with $\eta = 1.3$ mPa s in a microchannel 200 $\mu$m long and 20 $\mu$m wide and tall would have $\beta \approx 2$. These parameters correspond to Reynolds numbers ${\rm Re} \sim 100$.   

\begin{figure}[t]
\includegraphics[width=0.48\textwidth]{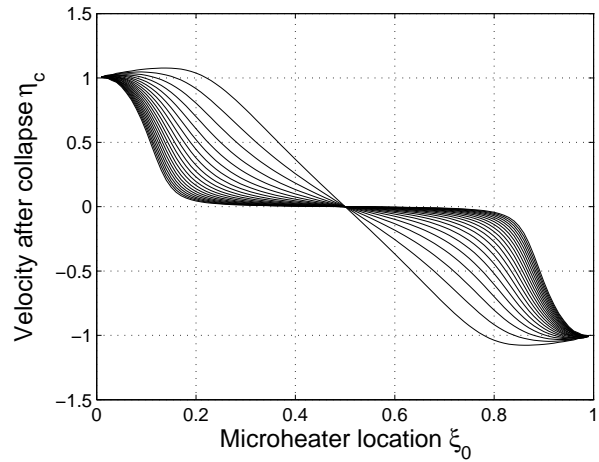}
\caption{Same as Fig.~\ref{oned:fig:twelve}, but the secondary pumping effect $\eta_c$ is shown.}   
\label{oned:fig:thirteen}
\end{figure}

\begin{figure}[t]
\includegraphics[width=0.48\textwidth]{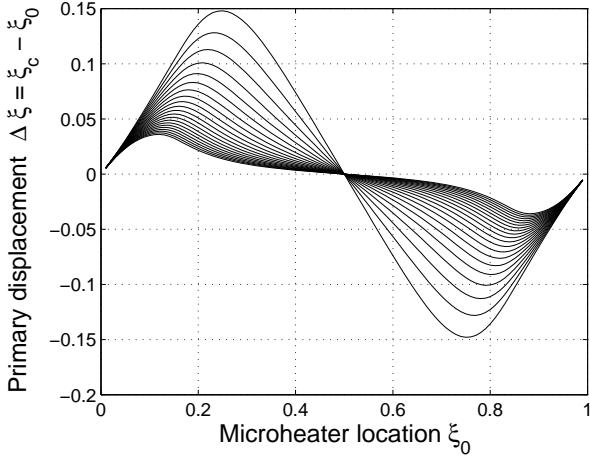}
\caption{Dimensionless primary displacement $\Delta \xi = \xi_c - \xi_0$ in the asymmetrical model $m = 1$ for $\gamma_1 = \gamma_2 = 1$ and different friction parameters $\beta = 0.0,0.5,1.0,\ldots,10.0$. The bubble strength is $\alpha = 0.5$. The trace with the largest amplitude corresponds to $\beta = 0.0$. The curves are obtained by numerical solution of the pump equations (\ref{onedf:eq:twelve}) and (\ref{onedf:eq:thirteen}).}   
\label{oned:fig:fourteen}
\end{figure}

\begin{figure}[t]
\includegraphics[width=0.48\textwidth]{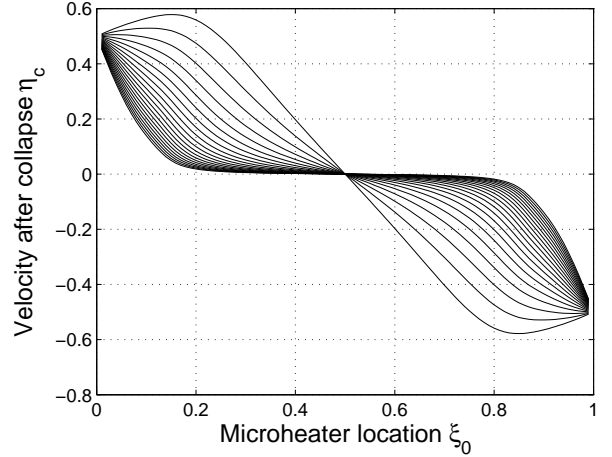}
\caption{Same as Fig.~\ref{oned:fig:fourteen}, but the secondary pumping effect $\eta_c$ is shown.}   
\label{oned:fig:fifteen}
\end{figure}

A nonzero $\beta$ introduces additional nonlinearities into the equation of motion. Although some useful results can still be obtained analytically (particularly, in the postcollapse phase; see Sec.~\ref{onedf:sec:five}), final pumping effects are difficult to derive. We therefore resort to numerical analysis. Results are presented in Figs.~\ref{oned:fig:twelve} and \ref{oned:fig:thirteen} for the symmetrical model and in Figs.~\ref{oned:fig:fourteen} and \ref{oned:fig:fifteen} for the asymmetrical model. In selecting a representative bubble strength, we chose $\alpha = 0.5$, which roughly corresponds to a vapor bubble pressure of $p_v =$ 8-10 atm (cf. caption to Table~\ref{onedf:tab:one}).   

As expected, the finite viscosity systematically reduces both inertial effects. However, the reduction is not uniform across all microheater locations. Reductions are stronger for $0.1 < \xi_0 < 0.9$, but are much less for the locations close to the channel ends, as can be observed in the plots. For the primary effect (Figs.~\ref{oned:fig:twelve} and \ref{oned:fig:fourteen}) such a nonuniform change results in a systematic shift of the optimal location closer to the end of the channel. For the secondary effect, the maximum is eliminated altogether. This behavior is another indication of the complexity of inertial pumping.  

Our results suggest that the inertial pump can operate down to ${\rm Re} \sim 10$ or even below, but its efficiency drops quickly. This is also consistent with the experimental data reported in Ref.~\cite{Torniainen2012}. Pumping through 1-$\mu$m-wide channels should still be possible; however, the microheater will have to be positioned very close to channel's end.

\section{\label{onedf:sec:five}
Post-collapse phase
}

For realistic conditions, most of the net flow happens in the postcollapse phase when fluid moves by inertia against friction forces. It is therefore important to know the details of the postcollapse flow. The equation of motion was derived in Sec.~\ref{oned:sec:twofour}, Eq.~(\ref{onedf:eq:twentyone}). Since this equation is simpler than its counterparts of the expansion-collapse cycle, it is possible to obtain analytical solutions even for nonzero frictions $\beta$ and pressure heads $\gamma_1 - \gamma_2$. In this section the most common case of a zero pressure head is solved. The more complex $\gamma_1 \neq \gamma_2$ case will not be considered herein.    

For the symmetrical model $m = 0$, the equation of motion reduces to
\begin{equation}
\xi'' + \beta \xi' = 0  \: , \hspace{0.5cm} \xi'(\tau_c) = \eta_c \: ,  
\label{oned:eq:onehtwenty}     
\end{equation}
with the obvious solution
\begin{eqnarray}
\xi'(\tau) & = & \eta_c \, e^{-\beta (\tau - \tau_c)} \: ,  
\label{oned:eq:onehtwentyone}     \\
\xi(\tau)  & = & \xi_c + \frac{\eta_c}{\beta} \left[ 1 - e^{-\beta (\tau-\tau_c) } \right] \: .
\label{oned:eq:onehtwentytwo}
\end{eqnarray}
Flow slows down according to a simple exponential law. Total displacement of the postcollapse phase is $\eta_c/\beta$. It diverges at zero viscosity. 

The initial position $\xi_c$ and velocity $\eta_c$ of the postcollapse phase are in general obtained from numerical solution. At small $\beta$, analytical expressions derived in Sec.~\ref{onedf:sec:threetwo} can be used. For example, at weak asymmetry $\xi_c$ and $\eta_c$ are given by Eqs.~(\ref{onedf:eq:thirtyseven}) and (\ref{oned:eq:eighty}) and at strong asymmetry by Eqs.~(\ref{oned:eq:eightyfive}) and (\ref{oned:eq:eightynine}), respectively. 

In the asymmetrical model $m = 1$, the Bernoulli term changes sign depending on flow direction. However, in the case of a zero pressure head, flow direction does not change with time. It is therefore sufficient to consider only one, for example, positive flow direction $\xi'(\tau) > 0$. The dynamic equation then reads   
\begin{equation}
\xi'' + \frac{1}{2} \, \xi'^2 + \beta \xi' = 0  \: .  
\label{oned:eq:onehtwentythree}     
\end{equation}
Both integrations are elementary, yielding
\begin{eqnarray}
\xi'(\tau) & = & \frac{2\beta \, \eta_c}{(\eta_c + 2\beta) \, e^{\beta(\tau-\tau_c)} - \eta_c } \: .  
\label{oned:eq:onehtwentyfour}     \\
\xi(\tau)  & = & \xi_c + 
2 \ln{ \left\{ \frac{\eta_c}{2\beta} \, 
\left[ 1 - e^{-\beta(\tau-\tau_c)} \right] + 1 \right\} } \: .  
\label{oned:eq:onehtwentyfive}     
\end{eqnarray}
The total displacement of postcollapse flow is finite
\begin{equation}
\Delta \xi(\infty) = 2 \ln{ \left( \frac{\eta_c}{2\beta} + 1 \right) }  .  
\label{oned:eq:onehtwentysix}     
\end{equation}
In the zero-viscosity limit, the velocity decays to zero as $2\eta_c/[2 + \eta_c(\tau-\tau_c)]$, but the overall displacement diverges logarithmically. Again, analytical formulas for $\xi_c$ and $\eta_c$ from Sec.~\ref{onedf:sec:threethree} can be used at small $\beta$ to estimate a total pumped volume.

\section{\label{onedf:sec:six}
Summary 
}

A micropump is required for any active fluid-handling system. Inertial micropumps do not contain moving parts and can be made in large quantities by batch fabrication processes. As such, they are an excellent candidate for the universal integrated pump of chip-scale fluidics. The physics behind the pump operation is defined by a subtle balance between the pressure, viscous, and inertial forces. A high-pressure vapor bubble generated by a microheater expels fluid from the channel to reservoirs, but after that flow reverses under a negative pressure difference. Because of a smaller mass and inertia the shorter arm reverses flow direction earlier and then has more time to accelerate during inflow. By the time of collision, the shorter arm acquires a larger velocity and momentum than the longer arm. The two columns collide at a point that is shifted from the initial point of expansion, which constitutes the primary pumping effect. A nonzero total momentum ensures postcollapse flow from the short toward the long side of the channel, which is the secondary pumping effect. Total net flow is the sum of the primary and secondary contributions.

In this paper the inertial pumping action has been studied within a simplified one-dimensional dynamic model. Transverse motion within the channel has been neglected and the entire dynamics reduced to that of a fluid-vapor interface treated as a mathematical point. Despite these simplifications, the one-dimensional model captures the main features correctly, while using scaled units reduces the number of independent parameters. The model becomes an efficient tool for analyzing different pumping regimes. A major challenge lies in understanding the pressure at the channel-reservoir interface. The transient nature of the flow, vortex formation in the reservoir, nonuniform velocity across the channel cross section during flow reversal, and other factors make the selection of a single boundary condition nontrivial. Until a full understanding is achieved via additional experimental and numerical work, the current approach is to assume a physically reasonable boundary condition and analyze its consequences. In this paper two possible choices have been studied in detail. In the simplest symmetrical model the interface pressure is assumed constant and set to be equal to the bulk reservoir pressure. In the more complex asymmetrical model, during inflow the interface pressure is reduced from the bulk value by a Bernoulli-type correction.      

A key finding of the paper is that both models contain pumping effects and their properties are similar for weak to intermediate bubble strengths (which is the most realistic situation). In particular, the two models predict an optimal microheater location in the 0.2-0.3 range for both the primary and secondary effects. A nonzero viscosity dampens net flow in both models in a similar fashion. The predictions begin to diverge for very strong bubbles $\alpha > 1$, as can be seen by comparing Figs.~\ref{oned:fig:six} and \ref{oned:fig:seven} with Figs.~\ref{oned:fig:ten} and \ref{oned:fig:eleven}. The approach adopted in this paper can be extended in a number of ways, including sequential firing, nonzero pressure heads, and branched channels.

\begin{acknowledgments}

The authors wish to thank Andrew Cihonski for useful discussions on the physics of inertial pumping and Tom Cooney, Mary Kent, Larry Mull, Michael Regan, and Susan Richards for supporting this work. 

\end{acknowledgments}

\appendix

\section{\label{onedf:sec:app:c}
Derivation of Eq.~(\ref{onedf:eq:six}). 
}

For Poiseuille flow in a cylindrical channel of radius $R$, the velocity profile and the total flow are given by
\begin{equation}
v_z(r) = \frac{\Delta p}{4 \eta L} \left( R^2 - r^2 \right) ,
\label{onedf:eq:appc:two} 
\end{equation}
\begin{equation}
W = \frac{\pi \Delta p}{8 \eta L} \, R^4 \: ,
\label{onedf:eq:appc:three} 
\end{equation}
where $\eta$ is the bulk viscosity of the fluid. Connection to the one-dimensional model studied in the main text is done by identifying cross-section average velocity $W/A$ with fluid velocity $\dot{x}$ of the one-dimensional model. Expressing here the pressure difference 
\begin{equation}
\frac{\Delta p}{\eta L} = \frac{8 \dot{x}}{R^2} \: ,
\label{onedf:eq:appc:four} 
\end{equation}
and substituting into Eq.~(\ref{onedf:eq:appc:two}), the velocity profile is rewritten as 
\begin{equation}
v_z(r) = 2 \dot{x} \left( 1  - \frac{r^2}{R^2} \right) \: .
\label{onedf:eq:appc:five} 
\end{equation}
The only nonzero component of the stress tensor is
\begin{equation}
\sigma_{zr} = \eta \, \frac{\partial v_z}{\partial r} = - 4 \eta \, \dot{x} \, \frac{r}{R^2}\: .
\label{onedf:eq:appc:one} 
\end{equation}
The total viscous force acting on a cylinder surface of length $x$ is 
\begin{equation}
F_{\rm visc} = \left. \sigma_{zr} \right\vert_{r = R} \cdot 2\pi R \, x 
= - 8 \pi \eta \, x \dot{x} \: .
\label{onedf:eq:appc:six} 
\end{equation}
A comparison with the definition of the viscous force term (\ref{onedf:eq:four}) and (\ref{onedf:eq:five}) yields the relationship (\ref{onedf:eq:six}). 

Notice that the dimensionless conversion factor $8 \pi \approx 25$ is unusually large. For example, a 30\% solution of glycerol in water at 45 $^{\circ}$C with $\eta = 1.3$ mPa s would have an effective friction coefficient of $\kappa \approx 32$ mPa s. In Ref.~\cite{Torniainen2012} we reported a value of $\kappa = 28$ mPa s extracted from fits to full-scale CFD simulations, which is close to the above estimate.

\providecommand{\noopsort}[1]{}\providecommand{\singleletter}[1]{#1}%

\end{document}